HAVE 2005 –  IEEE International Workshop on
Haptic Audio Visual Environments and their Applications
Ottawa, Ontario, Canada, 1-2 October 2005# Prop-Based Haptic Interaction with Co-location and Immersion: an Automotive Application

Michaël Ortega[1,2], Sabine Coquillart[2]
[1]PSA Peugeot Citroën
[2] i3D - INRIA Rhône-Alpes - GRAVIR
655, avenue de l'Europe, 38334 Montbonnot, France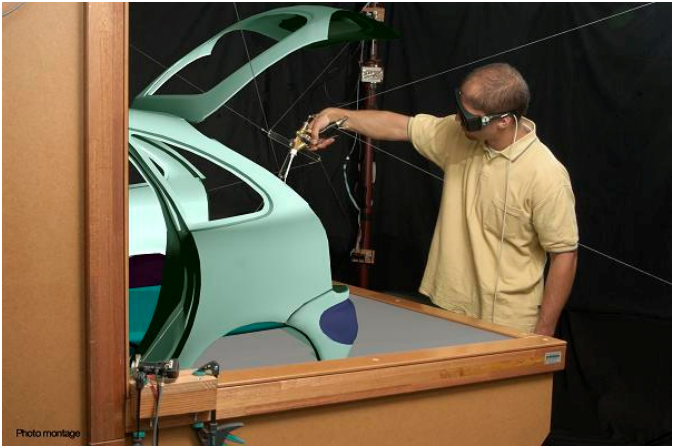

Figure 1: Putty application on a Citroën Picasso car

**Abstract** – *Most research on 3D user interfaces aims at providing only a single sensory modality. One challenge is to integrate several sensory modalities into a seamless system while preserving each modality's immersion and performance factors. This paper concerns manipulation tasks and proposes a visuo-haptic system integrating immersive visualization, tactile force and tactile feedback with co-location. An industrial application is presented.*

**Keywords** – *virtual reality, 6dof force feedback, prop, two-screen workbench, co-location.*## I.   INTRODUCTION

Most research on 3D user interfaces aims at providing only a single sensory modality. One challenge [1] is to integrate several sensory modalities into a seamless system while preserving each modality's immersion and performance factors. This paper concerns manipulation tasks and aims at providing a visuo-haptic system integrating immersive visualization, force and tactile feedback. Among the most important immersion and performance factors to be preserved, one can mention:

- large display and manipulation space
- stereovision
- head-tracking / co-location
- shadows
- 6dof force feedback
- props

Toward this goal, this paper proposes an extension of the Stringed Haptic Workbench [24] visuo-haptic configuration that integrates prop-based tactile feedback. This integration of immersive visualization, force and tactile feedbacks into a single system opens the door to new unexpected applications where immersion and interaction realism is critical. The system has been tested with an automotive putty application task. Informal user evaluations are presented. They highlight the benefits of the approach which is general enough to be applicable to other tasks and applications requiring realistic interaction with force and tactile feedback.

The next section of the paper presents previous work both on visuo-haptic immersive configurations and on tactile/grasp feedback. Section 3 describes the proposed prop-based visuo-haptic configuration. An industrial application and some aspects of its informal evaluation are presented in Section 4 and 5. Section 6 concludes the paper and proposes future work.

## II.   RELATED WORK

### A. Visuo-haptic VR configuration

Projection-based Virtual Environments such as CAVEs™ [9] or Workbenches [16], are the most popular VR configurations. They provide a large number of performance/immersion factors like stereoscopic visualization, large screens, large manipulation space, high resolution, head tracking, co-location, etc. However, adding force feedback to these configurations without degrading their performance/immersion factors is not an easy task. The main problem comes from co-location. In order to preserve co-location after the integration of haptics, the haptic system must preserve the VR configuration performance/immersion factors such as the size of the manipulation space, or stereoscopic visualization.

0-7803-9377-5/05/$20.00 ©2005 IEEE

Unfortunately, most general purpose haptic devices, like the PHANToM [18], have been conceived as a single sensory feedback device, and are often used with desktop visualization configurations. Most of the time, they are not able to adapt to VR configurations, leading to a degradation of some of the performance/immersion factors of the VR configuration. As an example, the manipulation space is much smaller than the workbench space. It is even worse with a CAVE™. In addition, projection-based VR configurations only allow for correct occlusions when real objects are in front of virtual ones. Consequently, parts of the haptic device (like the arm of the PHANToM) which are behind virtual objects of the scene lead to occlusion problems which may lead to a degradation of the stereoscopic effect.

Very few general purpose haptic systems have been integrated within large screen projection-based VR configurations. Both the University of Utah and North Carolina, Chapel Hill [2] [12] have installed a PHANToM on a one-screen workbench. The PHANToM is installed upside down and the haptic manipulation space is significantly reduced compared to the visual one. It would be even worse with a two-screen workbench. The PHANToM arm can also be in a position where it should be hidden by virtual objects and this situation leads to non-correct occlusions and disturbs stereovision.

Some authors have installed haptic systems such as the PHANToM or the Virtuose [11] either inside a CAVE™ or in front of large screens [3] [10] but to our knowledge, there are no attempts to preserve co-location if any.

To our knowledge, the only large screen projection based VR configurations equipped with a general purpose haptic system without loosing much in the manipulation space nor in occlusions are configurations equipped with the Spidar system [13] [4] [24]. The Spidar is a string-based haptic system which combines two critical advantages: it allows for large manipulation spaces and it is almost invisible (visually non intrusive). It is the reason why it has been installed with great benefits both on large screens and on a two-screen workbench. The workbench version is called "Stringed Haptic Workbench".

FishTank configurations [27] also often include haptics but most of the time, head-tracking is not provided and the co-location, if any, is supposed to be verified by the very low head movements. Finally, these configurations are limited by a relatively small manipulation space.

Other examples of haptic systems integrated within immersive projection-based configurations include configurations specific to an application, like a driving haptic simulator. They are out of the scope of this paper.

---

CAVE™ is a trademark of the university of Illinois

The work presented in this paper is an extension of the Stringed Haptic Workbench.

*B. Grasp feedback*

Realistic grasp feedback is also an important immersion/performance factor. Grasp feedback includes both tactile and shape feedback. Haptic systems exist for both but they are different and rarely integrated.
Tactile feedback includes different technologies such as temperature, air bubbles, needles matrices, etc. However, the technology is still quite immature and the realism not yet very high.
Special devices have been proposed for shape feedback. The most commons are exoskeleton hands [7] [25] [5]. Here again, the feedback is often not very realistic because it is quite partial (one point feedback for each finger instead of a continuous feedback on the whole hand).
Tactile devices are often not easy to combine with shape feedback devices. Exoskeleton hands require attachment of each finger of the user's hand to each exoskeleton finger which prevents a tactile device being touched with the finger. In addition, both are often difficult to integrate into immersive visuo-haptic configurations. Some of these devices are not portable and must be used with desktop configurations, or are visually invasive like exoskeleton hands and would perturb the visualization feedback of large screen visualization configurations with co-location mode.

The best known solution for providing a realistic grasp feedback consists of using props. Props are physical objects held in hand by the user. Props have been proposed for tasks such as application control [8], 3D objects manipulation [14] [23] and design. Several psychophysics experiments demonstrate the benefits of props [14]. Props provide stable grasp feedback, intuitive manipulation as well as realistic shape and texture rendering.

Some props include force feedback, such as car steering wheels [19] or joysticks, but most of the time, props do not provide force feedback. Props do not allow sensation of the collision with a surface touched by the prop itself. Combining props with force feedback is again a difficult task because most force feedback systems can't attach props in a flexible way. The PHANToM, for instance, only includes a stylus and a finger cap.
Lécuyer et al. [17] propose a system combining a prop and force feedback called HOMERE. They attach a white cane (prop) at the end point of an arm force feedback, the Virtuose. HOMERE is a navigation tool dedicated to blind persons. User experiments demonstrate the benefits of the system, but the system provides neither visualization nor co-location. It also provides only 3dof haptic feedback.

## III. THE PROP-BASED STRINGED HAPTIC WORKBENCH

Props are rarely integrated into immersive visual systems with co-location, large manipulation space, stereoscopic visualization and 6dof force feedback. This paper proposes to investigate the integration of props into the Stringed Haptic Workbench. Three major aspects of the integration are detailed:

- Attaching the prop to the force feedback system,
- Occlusions of the prop with the virtual model,
- Shadows of the prop onto the virtual model.

### A. Attaching the prop to the force feedback system

The original Stringed Haptic Workbench was limited to a 3dof Spidar haptic system. The version proposed here makes use of a 6dof version [15] which includes 8 motors positioned at the vertices of a hexahedron (see Figure 2). Proposing 6dof force feedback for object manipulation is critical in feeling torque. An additional advantage of 6dof compared to 3dof is the larger manipulation space. See [6] for more details.

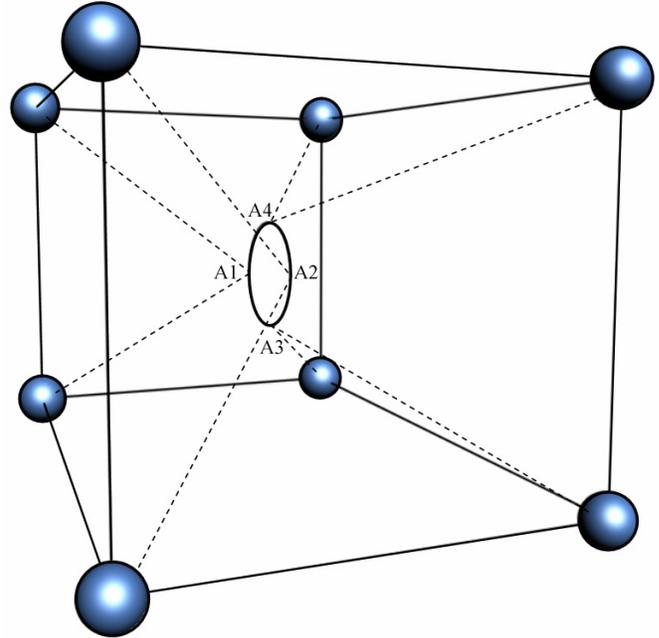

Figure 3: Spidar 8 strings fixation.

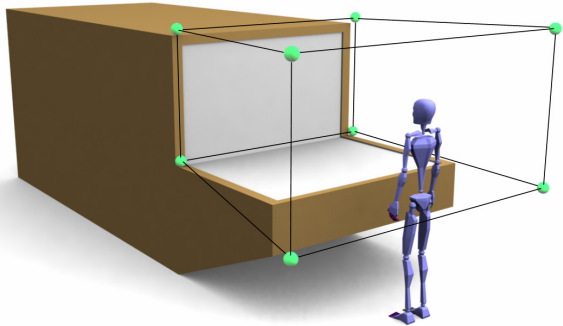

Figure 2: Position of 8 motors on the Stringed Haptic Workbench

One string is associated with each motor and has to be attached to the prop. In order to provide torque, the 8 strings are attached to 4 different points located in a circle as shown in Figure 3.

The choice of the diameter of the circle takes into account several parameters:

1. **Accuracy:** a large enough circle is required to ensure good accuracy and to avoid singularities [15]. A 10-30cm diameter seems to be a good range. We have used a 20cm diameter.

2. **Size of the prop:** the size of the circle must stay reasonable compared to the prop size. A circle which extends excessively far beyond the bounds of the prop could disturb both the visualization and the manipulation. It would also make the clamping of the strings onto the prop tricky. Thus, the size of the circle should stay in the range of the size of the prop, no more than doubling it.

If the size of the circle is in the range of the size of the object, and if the shape of the prop permits, one can attach the strings directly onto the prop. However this is most of the time not possible. In this case, we suggest attaching the strings onto a Plexiglas cross attached to the prop. Plexiglas has been chosen for its rigidity and transparency.

### C. Mixed props

Projection-based virtual environments do not allow virtual objects to occlude real ones. Props thus can't be moved behind a virtual object with correct occlusions. In order to solve this problem, mixed props are introduced. Mixed props consists of keeping as a physical prop only the part of the prop held in the hand and substituting the remainder of the prop by its virtual model. Mixed props provide several additional benefits:

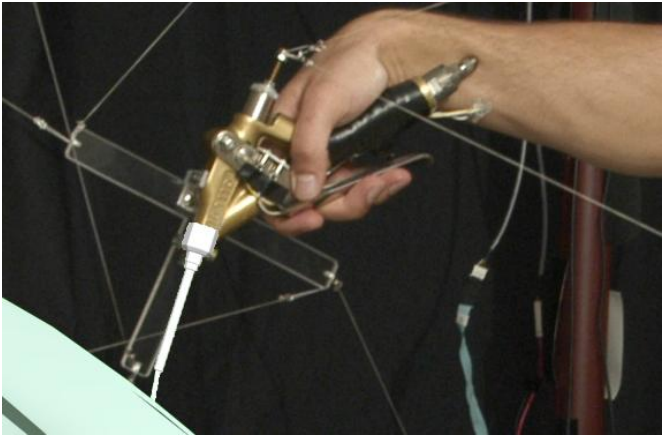

Figure 4: A putty gun with its virtual nose behind the car body.

- Mixed props can minimize the effect of calibration errors. Calibration errors can be characterized by a different positioning of the virtual prop (the model of the real prop used for computation, for instance, for collision detection), and the physical one (the one the user sees). This may for instance lead to collisions detected before the prop visually touches a virtual surface. If the part of the prop touching the surface is virtual, the collision appears when the user expects it from a visual point of view. However, the calibration problem doesn't disappear. It occurs at the junction between the virtual and the real parts of the prop. It may thus happen that these two parts appear to move independently of one another. However, as this part of the prop is usually not the main point of focus of the user, it is often not particularly disturbing.

- Substituting some parts of the prop with their virtual counterpart lead to a lighter physical prop. When the prop is too heavy compared to the force that the haptic system can return, it may happen that the reaction force is weakened. Using lighter physical props lowers this risk.

- Mixed props also allows the use of generic graspable parts together with more specific virtual parts which can easily be exchanged.

*D. Shadows*

Stereovision is only one of several solutions for providing depth information. Preliminary experiments on industrial cases have shown that stereovision and correct occlusions alones do not provide accurate enough depth information. It is difficult to evaluate the depth distance between two virtual objects, and consequently to anticipate collisions. Among the solutions known for providing depth information, one can mention stereo, occlusions, shadows, and accommodation. Because of the difficulty of evaluating depth using stereo and correct occlusions, and considering that the accomodation-convergence mismatch is unfortunatly an unsolved problem, it is proposed here to add shadows, known to improve depth perception [20] [26].

## IV. AUTOMOTIVE APPLICATION

The proposed immersive visuo-haptic system opens the door to new applications requiring a realistic integration of the three modalities mentioned above (visualization, force and tactile feedback). One such application, from the automotive industry, is described and evaluated in this section and the next one. It concerns putty application with a putty gun.

*A. Description of the application*

During the conception stage, car designers have to make sure that operators will easily be able to apply putty onto metallic junctions on the car body. Special attention has to be paid to three aspects:

- **Accessibility**. Accessibility evaluation,
- **Quality of the junction**. Evaluation of the quality of the junction where the putty has to be applied. Particular attention needs to be paid to the risk of having the putty gun slip off of the metallic seam, slowing down the assembly process.
- **Ergonomic**. Evaluation of the operator postures from an ergonomic point of view.

Until now, the only solution is to build a mockup of the car. The process is of course slow and expensive. A cheaper and faster solution consists in realizing the tests on virtual mockups. An additional advantage is that it can be done earlier in the conception process, which eases modifications. The remainder of this section presents this application in more detail.

*B. Hardware and Software Architecture*

For this application, the prop is a putty gun (see Figure 4). As described above, the Spidar is attached to the gun via a Plexiglas cross. The gun is treated as a mixed prop (see previous section). The physical part is the handle, while the nose is replaced by its virtual counterpart. In addition, a button has been added under the trigger for detecting when the user wants to lay down putty. The putty is simply visualized as an extrusion along the nose path.

Real-time shadows of the prop have been added. Figure 6 shows the shadow of both the nose and the physical graspable part of a putty gun. The shadow of the physical part is computed using a virtual replica which is not displayed to the user.

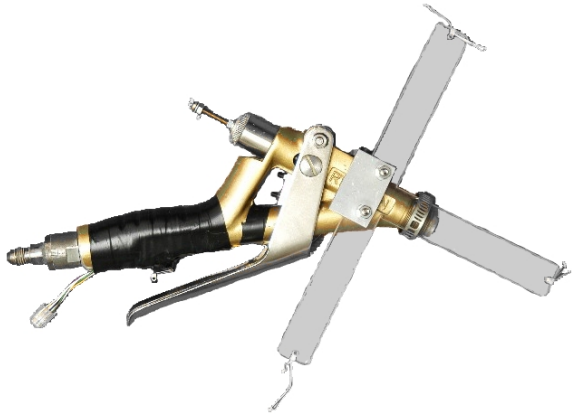

Figure 5: Plexiglas cross on a putty gun

The Spidar is connected to a Xeon 3.2Hz. On this PC, the application launches the dynamic engine loop (CONTACT Toolkit [21][22]) and the haptic controller loop. This computer communicates by UDP Protocols with a PC cluster. This cluster uses a proprietary OpenSG-based platform to manage visual display of the application, head-tracking and stereovision.

*C. Informal Evaluation*

The integrated solution presented in this paper has been informally evaluated with the industrial application presented in the previous section. As expected, the combination of immersive visualization, shadows, co-location, 6dof force-feedback and props representing the actual industrial tools greatly improves the realism of the interaction. User gestures are similar to real ones. PSA Peugeot Citroën representatives conducted informal studies. They applied virtual putty to a virtual car body and were able to determine critical regions. They unanimously approved the proposed solution and considered its industrial potential. Transfer of the proposed approach, including hardware and software, to PSA Peugeot Citroën is now under investigation.

## V. CONCLUSION AND FUTURE WORK

This paper proposes a solution for the integration of active force feedback and props in immersive visual display with co-location. A mixed (virtual/real) prop is attached to a non-intrusive haptic device, to provide realistic grasp information and 6dof force feedback. Special attention has been paid to preserving each modality's immersion and performance factors.

The proposed approach has been tested on an automotive industrial application. This first application has shown the potential of the approach, which is general enough to be applicable to other tasks and applications. Some of them are already under investigation, such as a screwing simulation with a pneumatic screw gun. Combining the proposed approach with two-hand manipulations is also planned for future work.

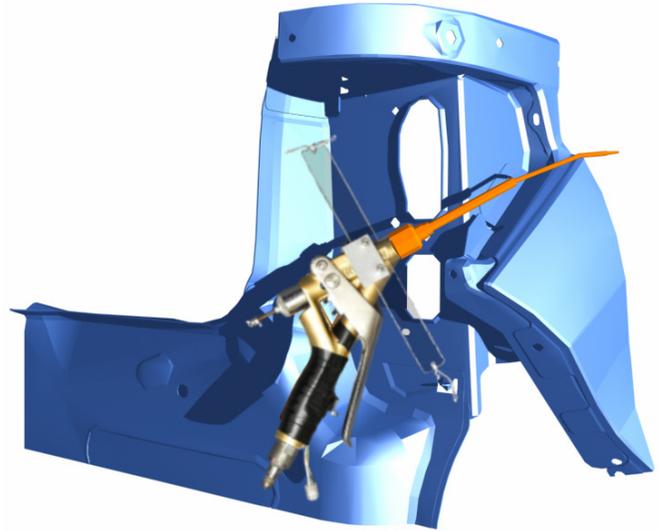

Figure 6: A putty gun with its virtual nose, casting a shadow on a car body.

## ACKNOWLEDGMENT

The authors would like to express their profound appreciation for the support and feedback from the PSA Peugeot Citroën representatives involved in the project. Thanks also go to Mathias Lorente for the conception of the proprietary VR platform.